\documentclass{aa}
\usepackage{graphicx,color}
\usepackage{float}
\usepackage{amsmath}
\usepackage{natbib}
\usepackage{subfigure}
\usepackage[breaklinks]{hyperref}
\bibpunct{(}{)}{;}{a}{}{,}

\begin{document}
% Title Page
\title{Instrumental oscillations in RHESSI count rates during solar flares}

\author{A. R. Inglis \inst{1,5} 
\and I. V. Zimovets \inst{2}
\and B. R. Dennis \inst{1}
\and E. P. Kontar \inst{3}
\and V. M. Nakariakov \inst{4,6}
\and A. B. Struminsky \inst{2}
\and A. K. Tolbert \inst{1,7}}
\institute{Solar Physics Laboratory, Heliophysics Science Division, NASA Goddard Space Flight Center, Greenbelt, MD, 20771, USA
\and Space Research Institute, Russian Academy of Sciences, Profsoyuznaya str. 84/32, Moscow, 117997, Russia
\and School of Physics and Astronomy, University of Glasgow, Glasgow, G12 8QQ, UK
\and Centre for Fusion, Space and Astrophysics, University of Warwick, Coventry, CV4 7AL, UK
\and School of Mathematics and Statistics, University of St Andrews, St Andrews, KY16 9SS, UK
\and Central Astronomical Observatory at Pulkovo of the Russian Academy of Sciences, 196140 St Petersburg, Russia
\and Wyle Information Systems Inc., McLean, VA 22102, USA}

\abstract {}{We seek to illustrate the analysis problems posed by RHESSI spacecraft motion by studying persistent instrumental oscillations found in the lightcurves measured by RHESSI's X-ray detectors in the 6-12 keV and 12-25 keV energy range during the decay phase of the flares of 2004 November 4 and 6.}
{The various motions of the RHESSI spacecraft which may contribute to the manifestation of oscillations are studied. The response of each detector in turn is also investigated.}
{We find that on 2004 November 6 the observed oscillations correspond to the nutation period of the RHESSI instrument.
These oscillations are also of greatest amplitude for detector 5, while in the lightcurves of many other detectors the oscillations are small or undetectable.
We also find that the variation in detector pointing is much larger during this flare than the counterexample of 2004 November 4.}
{Sufficiently large nutation motions of the RHESSI spacecraft lead to clearly observable oscillations in count rates, posing a significant hazard for data analysis. This issue is particularly problematic for detector 5 due to its design characteristics. Dynamic correction of the RHESSI counts, accounting for the livetime, data gaps, and the transmission of the bi-grid collimator of each detector, is required to overcome this issue. These corrections should be applied to all future oscillation studies.}

%This is supported by the absence of any oscillations in the data f\right] \right\rbrace \right\rangle rom other available instruments and analysis of RHESSI motions during the counterexample flare of 2004 November 4.}

\maketitle

\section{Introduction}
\label{intro}
Solar flare activity is often accompanied by quasi-periodic pulsations (QPP) in the flare emission, present over a wide range of electromagnetic wavelengths. QPP carry information about the physical processes operating in flares, and hence the study of oscillatory behaviour in solar events remains an active research topic, particularly in the context of coronal seismology \citep[see][for a recent review]{2009SSRv..149..119N}. In theory, the confirmed presence of such behaviour imposes additional constraints on the solar plasma which allow for the remote estimation of fundamental plasma parameters such as the density, temperature, and magnetic field strength, as well as insight into the basic mechanisms responsible for energy release, conversion, and particle acceleration.

Oscillatory behaviour is usually manifest in terms of the periodic modulation of solar emission associated with high-energy particles, and thus is observed directly using a number of relevant instruments. Principal among these are the Nobeyama Radioheliograph (NoRH), Yohkoh's Hard X-ray Telescope (HXT), and the Reuven Ramaty High Energy Solar Spectroscopic Imager (RHESSI). These instruments are preferred because they possess the necessary temporal resolution to resolve short-period, short duration events characteristic of the solar corona and solar flares. Moreover, these devices all have a spatial resolution of a few arcseconds, an important advantage.

Previous studies show that, in the majority of described cases, QPP are present in flaring lightcurves as short wave trains with varying periods and amplitudes
\citep[see, e.g.][]{Melnikov, 2008A&A...487.1147I, 2007AdSpR..39.1804N, 2009SSRv..149..119N, 2009SoPh..258...69Z}, although in a few cases harmonic oscillations of persistent amplitude and duration have been detected \citep[see, e.g.][]{1971Natur.234..140M, 2006SoPh..233...89K}. Moreover, QPP are often detected with several independent and spatially separated instruments, not only in hard X-rays and microwaves, but also at sub-millimeter wavelengths \citep{2009ApJ...697..420K, 2010ApJ...709L.127F} and in soft X-rays \citep[e.g.][]{2006ApJ...639..484M}. Hence the solar origin and nature of QPP in flares is well-established in general. However, in some cases, when the oscillatory pattern is seen with only a single instrument, there is a danger that it is an artifact of the observing instrument and not of solar origin.

RHESSI has contributed to a large number of oscillatory studies \citep[e.g.][]{2005A&A...440L..59F, 2006ApJ...644L.149O, 2006A&A...460..865M, 2008ApJ...684.1433F, 2008SoPh..247...77L, 2009A&A...493..259I, 2009SoPh..258...69Z, 2010SoPh..263..163Z, 2010ApJ...708L..47N} and is an excellent tool in this regard, due not only to its time cadence, but also to its wide coverage of the X-ray regime. RHESSI is capable of detecting emission from as low as 3 keV in soft X-rays, up to 17 MeV in gamma-rays. However, the RHESSI spacecraft is spinning at 15 rpm and is also subject to nutation and precession motions which are capable of causing misleading results.

In this paper, we examine instrumental oscillations in count rates detected by RHESSI in the decay phase of two large solar flares which occurred on 2004 November 4 and 6, respectively. The oscillations during the 2004, November 6 flare were previously misinterpreted as being of solar origin by \citet{ISI:000279654100009}. Using these examples, we illustrate the hazards involved in interpreting RHESSI data and how to produce X-ray lightcurves free of these instrumental effects.

\section{RHESSI}

RHESSI is a Sun-observing X-ray satellite launched in 2002 \citep{2002SoPh..210....3L}. With its nine germanium detectors, RHESSI records incoming photon flux over a wide range of X-ray energies, from 3 keV up to 17 MeV. These detectors are each fronted with collimating bi-grids of varying dimensions. Detector 1, with the finest grids, has a FWHM angular resolution of 2.3 arcseconds, while detector 9, with the coarsest grids, has a resolution of 180 arcseconds.

Each of RHESSI's detectors also has two moveable aluminium attenuators. These attenuators are activated at times of strong X-ray flux, such as during a powerful flare, and slide in front of the detectors to reduce the detected counts. This procedure helps to combat problems such as pulse pile-up, detector dead-time, and also avoids filling up the spacecraft's on-board memory too quickly.

Although primarily designed as an imager and spectrometer, RHESSI also provides excellent time series data for the study of solar flares. It is one of the few instruments in operation capable of providing solar hard X-ray data with sufficiently high time cadence to reliably study QPP.
Situated in low-Earth orbit, RHESSI experiences regular day-night cycles. A typical observation window for RHESSI lasts approximately 65 minutes, followed by roughly 30 minutes of eclipse, or night time. RHESSI's orbit also passes through the South Atlantic Anomaly (SAA). The collection of data is suspended while it passes through this region to avoid filling up the spacecraft memory.

\begin{figure}
 \begin{center}
  \includegraphics[width=8cm]{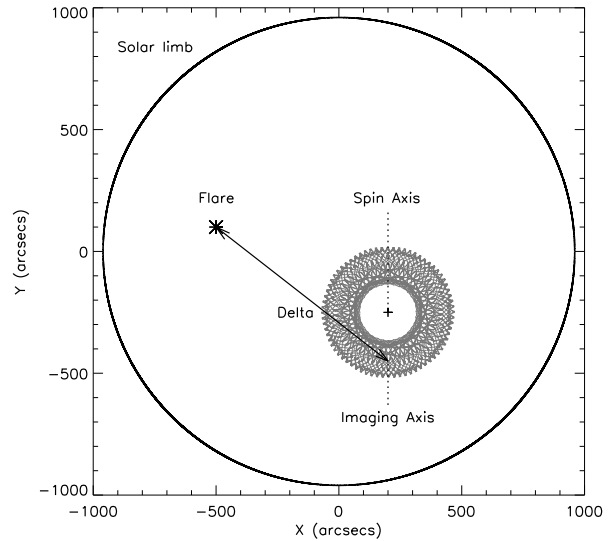}
  \caption{Illustration of the RHESSI satellite pointing behaviour. The imaging axis describes circles on the solar disk every 4 s, resulting in variations in $\delta$, the distance between the instantaneous telescope pointing and the X-ray source location.}
  \label{cartoon}
 \end{center}

\end{figure}

RHESSI's pointing motions can be described in terms of three main effects. Firstly, the satellite rotates on its axis approximately every 4 s. This rotation is essential to the imaging process, which relies on the bi-grid collimators. As the spacecraft rotates, the X-rays passing through the collimating bi-grids are modulated, allowing the X-ray source to be imaged \citep[see][for more information on the RHESSI imaging concept]{2002SoPh..210...61H}. Additionally, since the imaging axis and the spacecraft spin axis are slightly offset, the imaging axis describes circles on the solar disk around the spin axis every 4 seconds (see Figure \ref{cartoon}). The imaging axis is determined from the solar aspect system described by \citet{2002SoPh..210...87F}.

Since the satellite is spinning in free space it experiences nutation and precession, which produce motions of both the telescope imaging and spin axes. The amplitudes and phases of these periodic gyroscopic motions change with time as RHESSI is subject to small forces. The main force affecting these motions is from the Earth's magnetic field that is used through on-board magnetic torquer rods to both maintain the 15 rpm spin rate and follow the Sun as it appears to move at approximately 1 degree per day with respect to the background stars.

RHESSI is also affected by heating and cooling from the day-night cycle, which changes the length of the solar panels and other spacecraft components. Also, the motion of the attenuator plates as they are moved into and out of the detector field of view can induce temporary oscillatory motions of the imaging axis about the spin axis.

%Thirdly, the pointing axis of RHESSI experiences significant drift during each observation window. In general, the pointing centre at the beginning %of an observation is significantly towards solar west, and drifts towards the east during the observation interval. During night time, the pointing %system is deactivated, causing a drift back towards solar west during this eclipse window. 

Studies of oscillations and pulsations in X-ray data are invariably concerned with the analysis of time series data in order to determine whether any significant frequencies are present. Hence, all of the cyclical motions of RHESSI must be considered carefully. The 4 s rotation of RHESSI is commonly eliminated from any analysis by averaging count rates over a full rotation. Nutation and precession motions, however, are often overlooked or considered to be small effects.

By examining the time profiles of X-ray emission from two solar flares, we show that the nutation and precession motions of RHESSI can have a significant impact on the temporal analysis of count rate data. In the following sections we explore the reasons behind this, their observational manifestations, and the mitigating steps that can be taken to combat these effects.

\section{Modulation in the 2004, November 6 flare}

\begin{figure}
 \begin{center}
 \includegraphics[width=7.5cm]{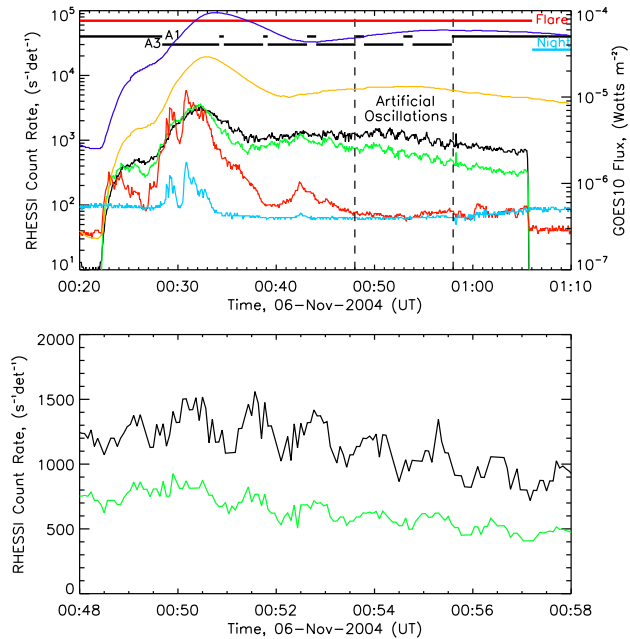} 
\caption{Count rates obtained as a function of time by RHESSI on 2004 November 6 at 6-12 keV (black), 12-25 keV (green), 25-50 keV (red) and 50 - 100 keV (turquoise), using front and rear segments of detectors 1, 3, 4, 5, 6 and 9. These count rates have been adjusted to compensate for attenuator state changes. Count rates observed by GOES-10 at 1 - 8 \r{A} and 0.5 - 4 \r{A} are marked by blue and orange, respectively. The lower plot shows an enlargement of the region exhibiting oscillations in the 6-12 keV (black) and 12-25 keV (green) bands, between 00:48 and 00:58 UT.}
\label{figure1}
 \end{center}

\end{figure}

Analysing the decay phase of the M9.3 GOES-class solar flare on 2004 November 6, harmonic oscillations of RHESSI count rates in the 6-12 and 12-25 keV energy channels were found \citep{ISI:000279654100009}. The flare itself was located in NOAA Active Region 10696, centred at approximately (-80, 80) arcseconds on the solar disk. The observed oscillations had a characteristic period of 75 s and are clearly visible in the time profile between 00:48 UT and 00:58 UT (see Figure \ref{figure1}).

 As stated in Section \ref{intro}, persistent QPP exhibiting such a consistent period over a duration of at least 10 minutes are not typical or likely during flares. Hence, although these observations were interpreted as manifestations of a magnetohydrodynamic process by \citet{ISI:000279654100009}, they are in fact the result of an instrumental issue. 

%With all solar observational studies it is desirable to have supporting, complementary information from different instruments, in order to reduce or eliminate the possibility of %an instrumental effect compromising the results. The flare of 2004 November 6 was also observed by the Nobeyama Radioheliograph in the microwave regime. In a number of previous %studies oscillations have been seen to occur simultaneously in X-rays and in microwaves
%\citep[e.g.][]{2001ApJ...562L.103A,2008A&A...487.1147I,2009A&A...493..259I, 2009SoPh..258...69Z}. Hence, examination of Nobeyama Radioheliograph data provides a useful - though %not conclusive - test. In Figure \ref{norh_flux} the lightcurve from NoRH observing at 17 GHz is shown.

%\begin{figure}
% \begin{center}
 % \includegraphics[width=8cm]{6_nov_04_17ghz_v2.ps}
%  \includegraphics[width=9cm]{norh_figure.ps} 
% \caption{The flare of 2004 November 6 as observed by the Nobeyama Radioheliograph at 17 GHz, co-temporal with the RHESSI observations shown opposite.}
%\label{norh_flux}
% \end{center}

%\end{figure}

%Although the emission at the flare onset (around 00:30 UT) corresponds well with the observed X-ray time profile, Figure \ref{norh_flux} shows that the oscillations seen in the %X-ray regime during the decay phase of the flare are not reproduced in the microwave range. This in itself does not invalidate the X-ray oscillations, since it is quite possible %for X-ray emission to occur independently of radio emission. However the lack of corroboration is suggestive of a problem with the X-ray data. 

Key to understanding this is the behaviour of the RHESSI satellite during this time. Since the satellite not only rotates, but also nutates and precesses, the average transmission of X-rays through the detector grids will change as a function of time. If these motions are a significant fraction of the field of view through the grids, they can have a noticeable effect on the observed count rates.

\begin{figure}
 \begin{center}
   \includegraphics[width=9cm]{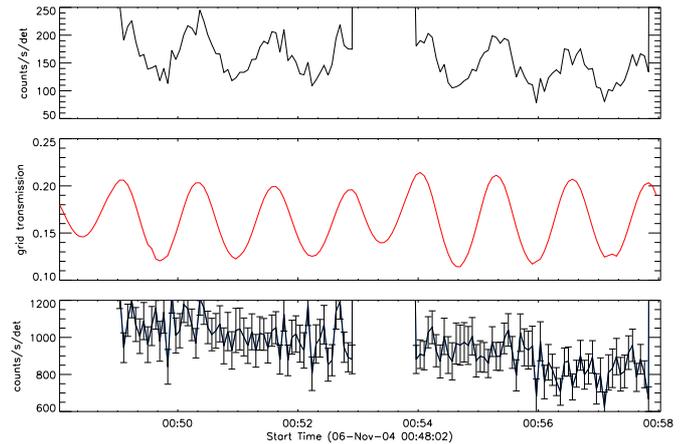}
   \caption{Top: RHESSI count rates detected on 2004, November 6 in the 6-12 keV range using detector 5, corrected for livetime only. Center: Grid transmission coefficients obtained for detector 5 as a function of time. Bottom: RHESSI count rates in the 6-12 keV range using detector 5, corrected for livetime and grid transmission. The error bars show the uncertainty due to Poisson statistics.}
\label{fix}
 \end{center}

\end{figure}

The flare of 2004, November 6 is an example of a case where this occurs. To illustrate this we show the modulation of the count rates in detector 5 during the decay phase between 00:48 UT and 00:58 UT, along with the grid transmission fraction averaged over the 4 s time resolution of the lightcurve (see Figure \ref{fix}). It is immediately clear from this figure that the changes in grid transmission have resulted in the appearance of the oscillations in the lightcurves (Figure \ref{figure1}). The count rates corrected dynamically for grid transmission are shown in the bottom panel of Figure \ref{fix}.

To account for these nutation motions it is necessary to
dynamically correct the observed counts for both livetime and grid
transmission, a feature which is not currently included in the RHESSI software for the generation of lightcurves or spectra. OSPEX\footnote{\url{http://hesperia.gsfc.nasa.gov/ssw/
packages/spex/doc/ospex_explanation.htm}} (a spectral data analysis package), for example, is generally used with a single grid transmission correction matrix to the selected time interval, which usually includes the complete flare.

The dynamic correction for livetime and grid transmission can be
obtained from the calibrated event list, a RHESSI data object \citep[see][for more information on RHESSI software design]{2002SoPh..210..165S}. This object summarises the state of RHESSI at any given time and contains information on counts, aspect solution, livetime, and data gaps. Hence, the procedure is to obtain the counts, livetime and grid transmission
for short time bins (typically 0.1 s) using the calibrated event list, and subsequently
divide the number of counts recorded in each time bin by the livetime and grid transmission. The resulting
count rates should be free of any instrumental oscillations on timescales
of seconds or minutes. An example of this procedure may be found in \verb,hsi_fluxvar_example.pro,, within the \verb,hessi, branch of SolarSoftWare (SSW).

%We illustrate this procedure in Figure \ref{fix}. Here the top panel shows the large modulation in detector 5 count rates before the grid transmission correction. The center panel shows the grid transmission coefficient as a function of time. The bottom panel shows the resulting count rates when the grid transmission correction is applied. 

 %The 75 s modulation has clearly been removed; the error bars illustrate that the final result is a smooth decay with random Poisson fluctuations.

%One complication is that data gaps are not accounted for using this method, which may cause fluctuations not accounted for by the error bars. This is primarily an issue for the %coarse grids of detectors 8 and 9 and is beyond the scope of this paper. 

%The grid transmission information may be obtained for any detector, or combination thereof, as a function of time. Hence, the artificial oscillations caused by nutation may be %removed from RHESSI count rates. 

Currently, these corrections are subject to certain limitations. Background subtraction, for example, is not included, which becomes an important factor during flares with a low signal-to-noise ratio. Also, the effects of data gaps are currently only accounted for on an average basis. For the detectors behind the coarsest grids (i.e. 7, 8 and 9) the collimator modulation period can be longer than the length of a data gap. In this situation, the measured rate in the detector depends on the phase of the data gaps in the modulation cycle, which itself depends on the source location. Hence, producing fully corrected lightcurves for these detectors is more difficult. However, the correction presented here is sufficient for detectors 1-6, which possess the finer grids.

In the following sections, we explore in more detail why these oscillations are manifest and examine their prevalence in RHESSI data.

\subsection{RHESSI pointing variation}

\begin{figure}
 \begin{center}
   \includegraphics[width=9cm]{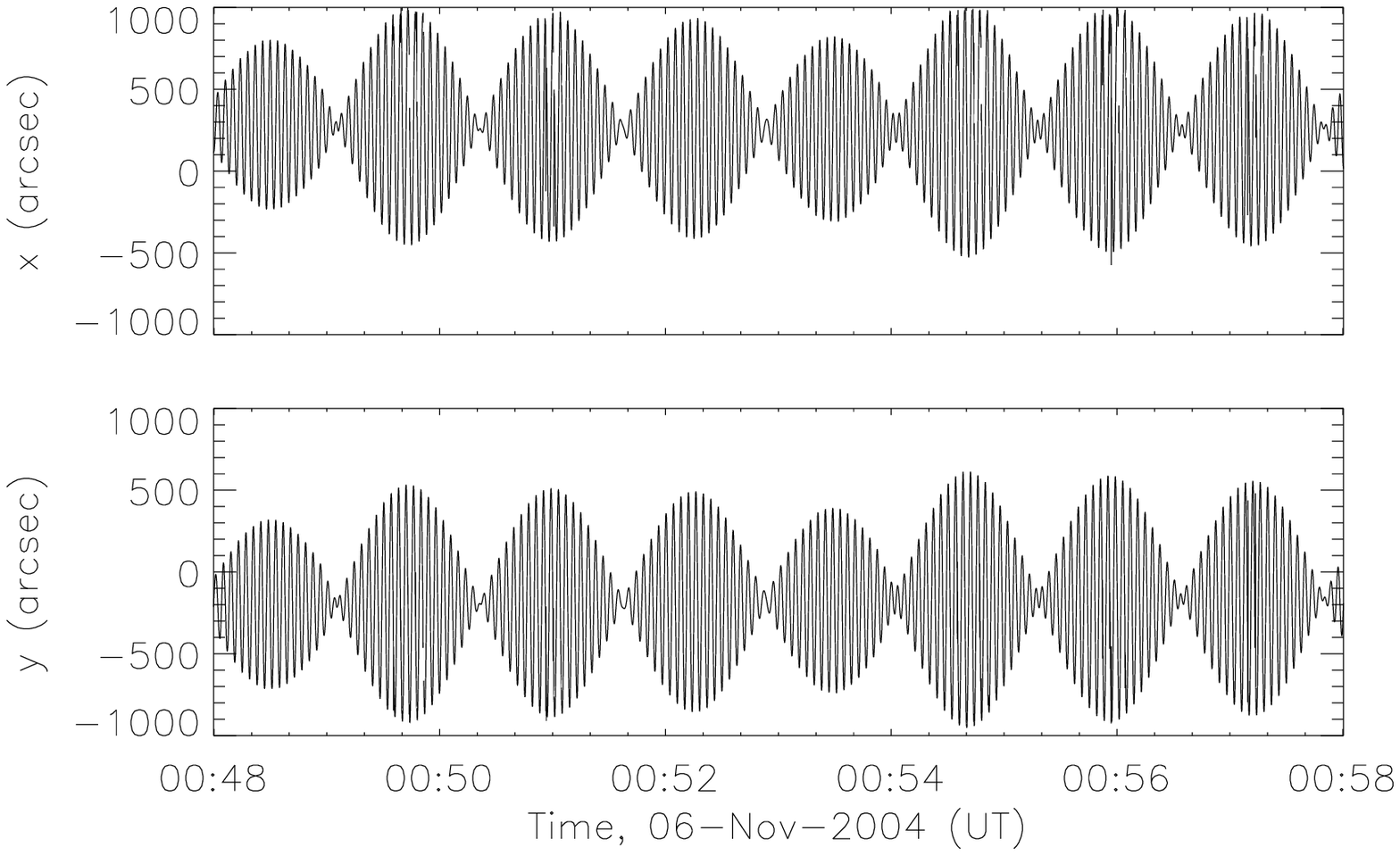}
\includegraphics[width=8cm,angle=90]{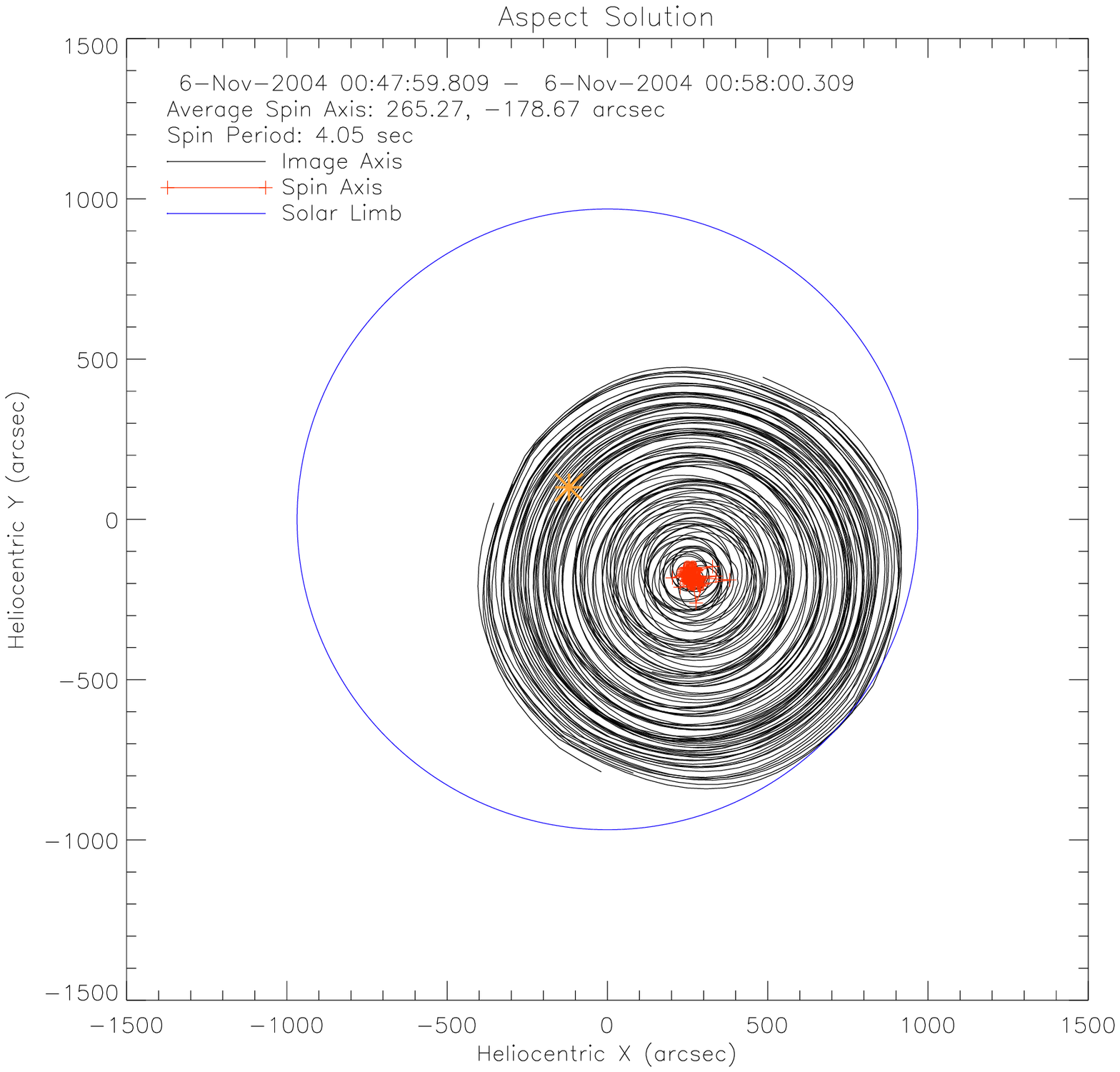}
 \caption{Top: RHESSI imaging axis pointing variations in $x$ and $y$ during the flare of 2004 November 6, showing the 4 s spin period and the 75 s nutation period. Bottom: RHESSI aspect solution, showing the varying location of the imaging axis (black) and spin axis (red) between 00:48 and 00:58 UT. The flare location is shown by the asterisk in yellow.}
\label{hessi_pointing}
 \end{center}

\end{figure}

The first step is to examine the behaviour of the RHESSI satellite during the flare of 2004 November 6. Since the period of the observed oscillations is in the vicinity of the RHESSI nutation period, we investigate the pointing variation of RHESSI during this time. Hence, in Figure \ref{hessi_pointing} (top panel) we plot the instantaneous location of the imaging axis in $x$ and $y$ as a function of time during the period when the oscillations were seen, 00:48 - 00:58 UT. Also in Figure \ref{hessi_pointing} (bottom panel) we plot the RHESSI aspect solution, showing graphically how the imaging axis and spin axis move during the same interval. 

It can be seen that, during this time, the pointing of RHESSI varies widely, both with the 4 s rotation period and with a secondary characteristic period of 75 s, corresponding to the period of the oscillations observed in Figure \ref{figure1}. The combined effect is that the imaging axis describes circles of varying radius around the spin axis, with this radius varying over a 75 s period. 

\begin{figure}
 \begin{center}
 \includegraphics[width=7.5cm]{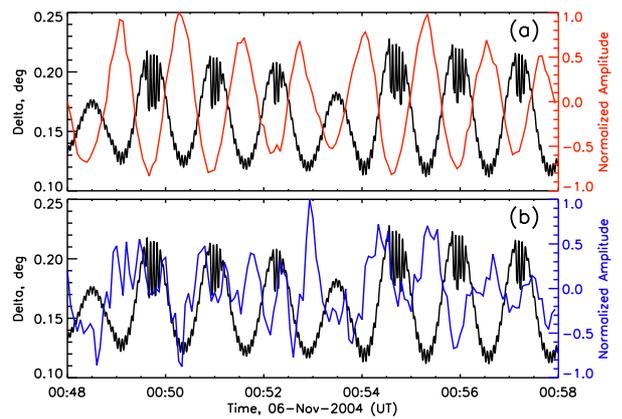} 
\caption{Variations in the distance between the flare source position and telescope position (black), and variations in smoothed, detrended, and normalized RHESSI count rates from detectors 5 (top panel, red) and 8 (bottom panel, blue) at 6-12 keV. Note that the two profiles are normalized independently and are in reality of different amplitudes (see Figure \ref{individual_detectors}).}
\label{d5d8}
 \end{center}

\end{figure}

In Figure \ref{d5d8} we plot the time profile of the 4 second averaged angular distance $\delta$ between the RHESSI imaging axis and the flare position on the Sun, shown as the black line. At the same time we plot for clarity the smoothed, detrended, and normalized profiles of the 6-12 keV count rates (called Normalized Amplitude for brevity) from detector 5 (red line, top panel) and detector 8 (blue line, bottom panel). Normalized Amplitude $NA\left(t\right)$ is defined as 
\begin{equation}
NA\left(t\right) = \frac{I_{20}\left(t\right)-I_{120}\left(t\right)}{max\left[I_{20}\left(t\right)-I_{120}\left(t\right)\right]},
\end{equation}
where $I_{20}\left(t\right)$ and $I_{120}\left(t\right)$ are the 4-s count rates during the interval 00:48 UT - 00:58 UT smoothed with a boxcar of 20 and 120 s, respectively. The 4-s count rates have been pre-adjusted to compensate for attenuator state changes. It is clearly seen, in detector 5 at least, that $\delta$ anti-correlates with Normalized Amplitude (or, equivalently, with the count rates). This means that the greater the angular distance between the imaging axis and the flare position, the smaller the detected X-ray flux. The situation in detector 8 however is not so clear.

Also visible in Figure \ref{d5d8} is some fine structure in the time profile of $\delta$ during some of the maxima. This is due to the imaging axis location being far (approximately 15 arcminutes) from the Sun center. The result is occasional faulty data where the solar aspect solution does not determine the location of the imaging axis correctly. 

% Despite the fact that the nominal 1-degree FOV of the RHESSI detector no. 5 is significantly larger than $\delta$, a modulation of the count rates is still present. 

%In Fig. 4 we plot time profile of the 4-second averaged angular distance $\delta$ between the RHESSI imaging axis and the flare position on the Sun, %jointly with the 20-second smoothed, detrended, and normalized corrected count rate (Normalized Amplitude) in the 6-12 keV energy channel integrated %over the RHESSI detectors No 1,3,4,5,6,9 during the time interval of the "pronounced" oscillations (see Fig. 2).

\begin{figure}
 \begin{center}
  \includegraphics[width=9cm]{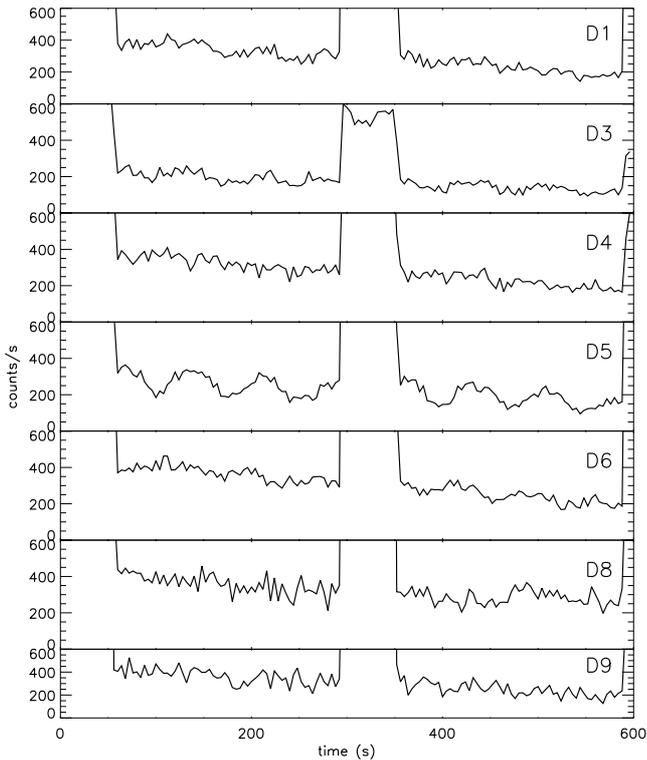}
  \caption{Counts detected by individual RHESSI detectors during the decay phase of the 2004 November 6 flare in the 6-12 keV energy range. The start time is 00:48 UT. The discontinuities in count rates are the result of the attenuator state changing between A3 and A1.}
\label{individual_detectors}
 \end{center}

\end{figure}

An important consideration is the differing responses of individual RHESSI detectors during a flare. Accordingly, we plot the counts detected by each of the RHESSI detectors, except detectors 2 and 7 \citep[see][for details of anomalies associated with these detectors]{2002SoPh..210...33S} in Figure \ref{individual_detectors}. Here the discontinuities seen in the count rates are the result of changes in attenuator states. It is evident that detector 5 experiences by far the greatest modulation, whereas in most of the other detectors the modulation of counts is either small or undetectable. The estimated modulation amplitude of each detector is summarised in Table \ref{table}. One difference between RHESSI's detectors is that they do not all have the same field of view through their bi-grid collimators. Detector 5, for example, has a FWHM field of view of less than 1 degree, whereas for detector 8 the nominal field of view is over 5 degrees. However, detectors 1-6 all possess comparable fields of view of 0.7 - 0.9 degrees, hence this alone does not explain the observations. 

%\textbf{It is useful to examine the modulation amplitudes of each detectors 1,3,4,5,6,8 and 9 in turn. Briefly examining the lightcurves of each detector, we find approximate %peak-to-valley amplitudes of $D1 \approx 40$ counts/s, $D3 \approx 40$ counts/s, $D4 \approx 60$ counts/s, $D5 \approx 120$ counts/s, $D6 \approx 50$ counts/s, $D8 < 40$ counts/s %and $D9 \approx 60$ counts/s.} 

The reason for the anomalous modulation amplitude of detector 5 lies in the pointing offsets of the subcollimators themselves. The grids in front of each RHESSI detector are mounted with collimating axes slightly tilted from that of the imaging axis (see Table \ref{table} for pre-launch data). These tilt angles are by far the greatest for detector 5. This is likely because the grids of detectors 5 - 9 are designed differently from those on detectors 1 - 4. Detector 5, possessing the finest grids of the van Beek design \citep[see][for more details of the grid manufacture]{2002SoPh..210....3L}, was the most difficult to manufacture. The result is that, during RHESSI's nutation and precession motions, the change in $\delta$, the distance between the imaging axis and the flare source, has a much greater effect on detectors where the offset angle is large. This modulation effect can be so strong that it is visible in the lightcurve even when the summation of counts is performed over all detectors.

\begin{table*}
\caption{RHESSI pre-launch detector grid offsets from the imaging axis and detector fields-of-view.}
\label{table}
\centering
 \begin{tabular} {c|c|c|c|c}
  \hline 
  \hline
  Detector & Front grid tilt (arcsec.) & Rear grid tilt (arcsec.) & FWHM FOV (deg.) & Modulation amplitude (counts/s) \\ [0.5ex]
  \hline
  1 & 75.6 & -21.6 & 0.9 & 40 \\ [0.5ex]
  2 & 57.6 & -36.0 & 0.8 & - \\  [0.5ex]
  3 & 90.0 & 64.8 & 0.8 & 40 \\ [0.5ex]
  4 & 79.2 & 36.0 & 0.8 & 60 \\ [0.5ex]
  5 & 507.6 & 482.4 & 0.7 & 120 \\ [0.5ex]
  6 & 324.0 & 111.6 & 0.7 & 50 \\ [0.5ex]
  7 & 118.8 & -57.6 & 3.2 & - \\ [0.5ex]
  8 & 79.2 & 147.6 & 5.4 & $<$40 \\ [0.5ex]
  9 & -133.2 & -180.0 & 2.0 & 60 \\ [0.5ex]
 \end{tabular}
\tablefoot{Detector information based on RHESSI grid parameter table: \\ \url{http://hesperia.gsfc.nasa.gov/rhessidatacenter/instrument/GPT3-7.htm}.}
\end{table*}

 It is clear that the observed modulation in count rates is a direct result of RHESSI's nutation and precession motions. However, the prevalence of this effect is not yet clear. The magnitude of RHESSI's motions varies significantly between times and events, and as we show in the following section, this determines whether observable 75 s oscillations appear in the lightcurves.

\subsection{A counterexample: the flare of 2004, November 4}

\begin{figure}
 \begin{center}
  \includegraphics[width=7.5cm]{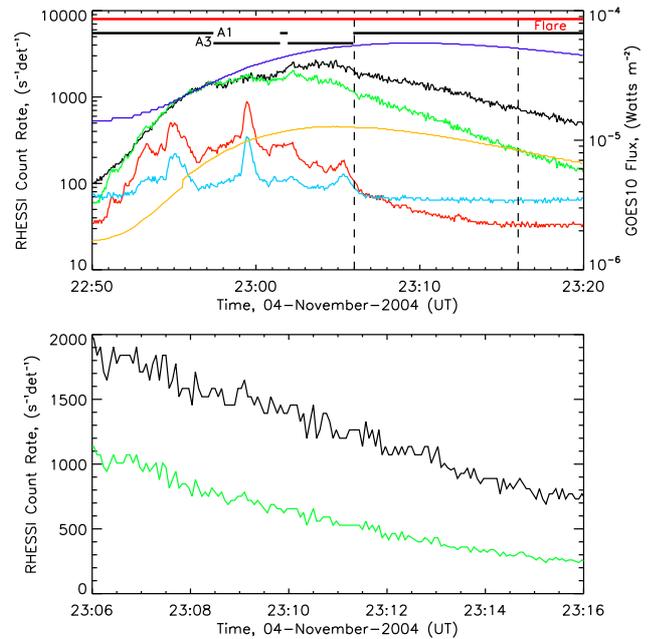}  
\caption{Count rates obtained as a function of time by RHESSI on 2004 November 4 at 6-12 keV (black), 12-25 keV (green), 25-50 keV (red) and 50 - 100 keV (turquoise), using detectors 1, 3, 4, 5, 6 and 9. These count rates have been adjusted to compensate for attenuator state changes. Count rates observed by GOES-10 at 1 - 8 \r{A} and 0.5 - 4 \r{A} are marked by blue and orange, respectively.}
 \label{04_nov_04_summary}
\end{center}

\end{figure}

In this section, we present analogous time series data of the M5.4 flare from November 4, 2004, the decay phase of which behaved in the standard way without showing any obvious oscillations of the RHESSI count rates (Figure \ref{04_nov_04_summary}). This flare occurred in the same active region as the flare of November 6, just over 25 hours earlier, located on the solar disk at approximately (-280, 80) arcseconds. Hence, we observe a similar strength flare with a similar source location that does not exhibit any signs of the oscillations described above. 

\begin{figure}
 \begin{center}
  \includegraphics[width=9cm]{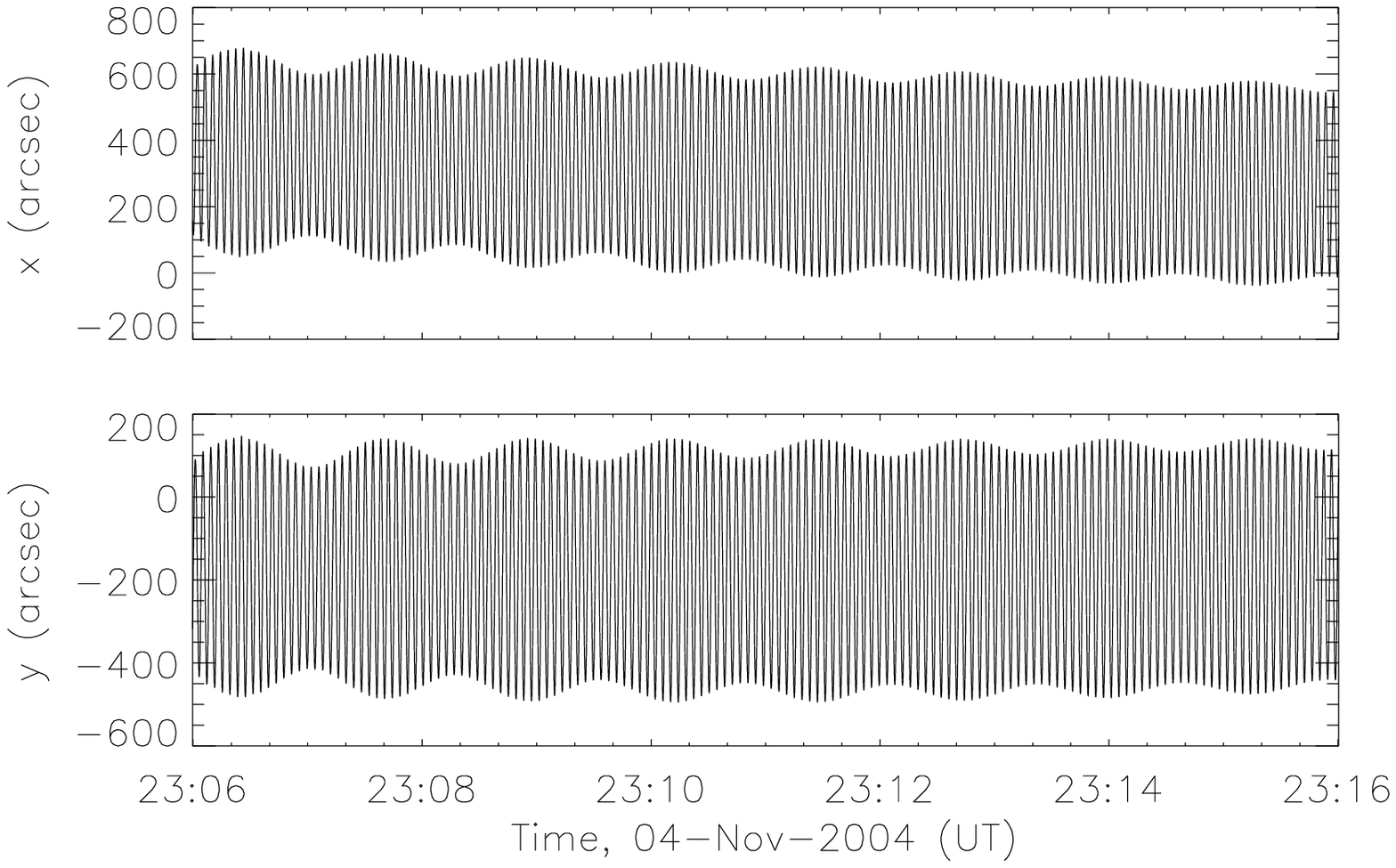}
 \includegraphics[width=8cm,angle=90]{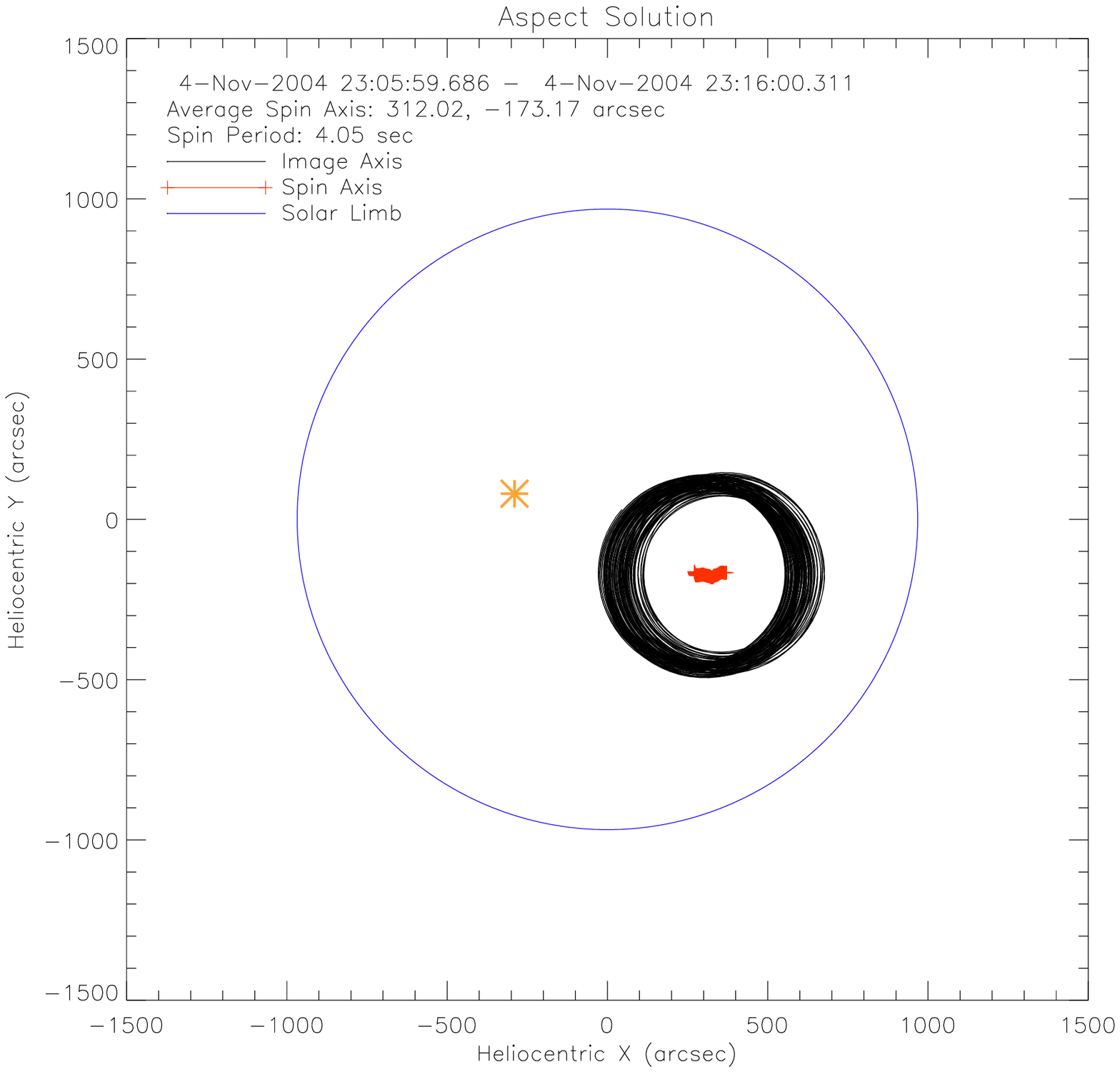}
\caption{Top: RHESSI pointing variations during the flare of 2004 November 4, showing the 4 s spin period and the 75 s nutation period. Bottom: RHESSI aspect solution, showing the varying location of the imaging axis and spin axis between 23:06 and 23:16 UT. The flare position is shown by the asterisk in yellow.}
 \label{04_nov_04_aspect}
\end{center}

\end{figure}

Investigating the behaviour of RHESSI during this time, it becomes apparent that the changes in the imaging axis location and in $\delta$ are much smaller on 2004 November 4 than for the 2004 November 6 flare, as shown in Figure \ref{04_nov_04_aspect}. Thus, as expected, there is a direct link between the magnitude of changes in $\delta$ and the appearance of artificial oscillations. 

%\begin{figure}
% \begin{center}
%  \includegraphics[width=0.47\textwidth, bb=14 5 480 270, clip=]{pointing_041104.ps}
%  \caption{Variations in the distance between the flare source position and imaging axis position (black) for the flare of 2004 November 4, and variations in smoothed, detrended, %and normalized RHESSI count rates from detectors 1, 3, 4, 5, 6, 8 and 9 in the 6-12 keV energy range (red).}
%\label{hessi_pointing_4_nov_04}
% \end{center}

%\end{figure}

\begin{figure}
 \begin{center}
  \includegraphics[width=9cm]{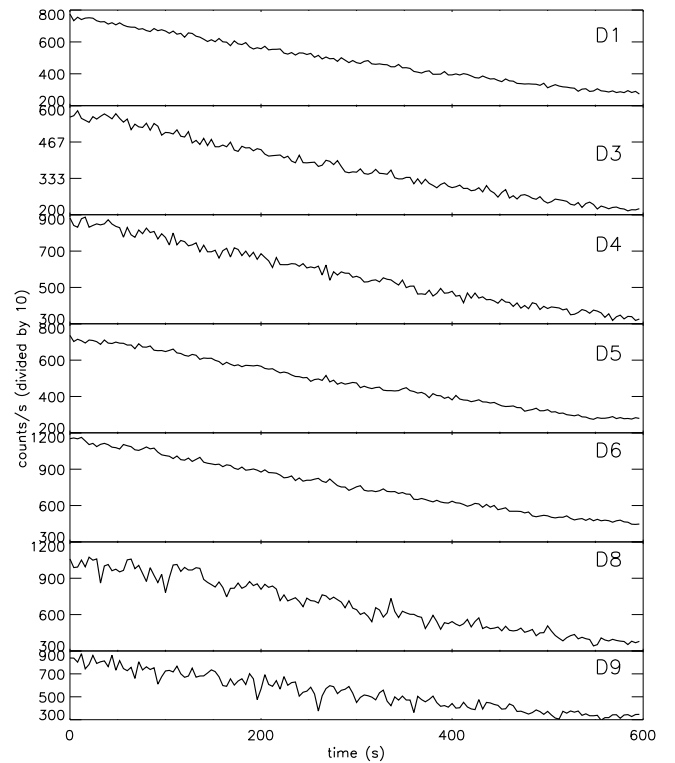}
  \caption{Count rates detected by individual RHESSI detectors during the decay phase of the 2004 November 4 flare in the 6-12 keV energy range. The start time is 23:06 UT.}
\label{detectors_04_nov_04}
 \end{center}

\end{figure}

Repeating our procedure from the previous section, we examine the counts detected in each RHESSI detector individually (see Figure \ref{detectors_04_nov_04}). In contrast with the previous flare, this event shows no significant modulation of the count rates due to RHESSI motions for any of the detectors. This is even true for detector 5, the source of anomalously large modulations during the flare of 2004 November 6.

Both flares are of GOES M-class and originate from the same active region on the solar disk. Hence, the principal difference between these two events, regarding the problem of oscillations, is the pointing behaviour of RHESSI. In the later flare of 2004 November 6 the pointing variation of RHESSI is rather large, leading directly to significant changes in $\delta$ as a function of time. This is markedly different from circumstances during the 2004 November 4 flare, where RHESSI's pointing remains relatively stable, resulting in only small changes in $\delta$. The result is that the oscillatory modulation of the detected counts does not appear in the latter case, or at least is of a much lower amplitude.

\subsection{Attenuator state changes}

One other difference between the flares of 2004 November 4 and 2004 November 6 is the dominant attenuator state during the decay phase. In the former case the decay phase of the flare is observed entirely in the A1 attenuator state. In the latter case, however, RHESSI was primarily operating in the A3 attenuator state, with regular brief changes to the A1 state to assess count rate levels.

However, during the impulsive phase of the 2004 November 4 flare, RHESSI was operating in the A3 attenuator state, before reverting to the A1 state at the onset of the decay phase at approximately 23:06 UT. Analysis of RHESSI pointing during this time is revealing.

\begin{figure}
 \begin{center}
\includegraphics[width=6cm, angle=90]{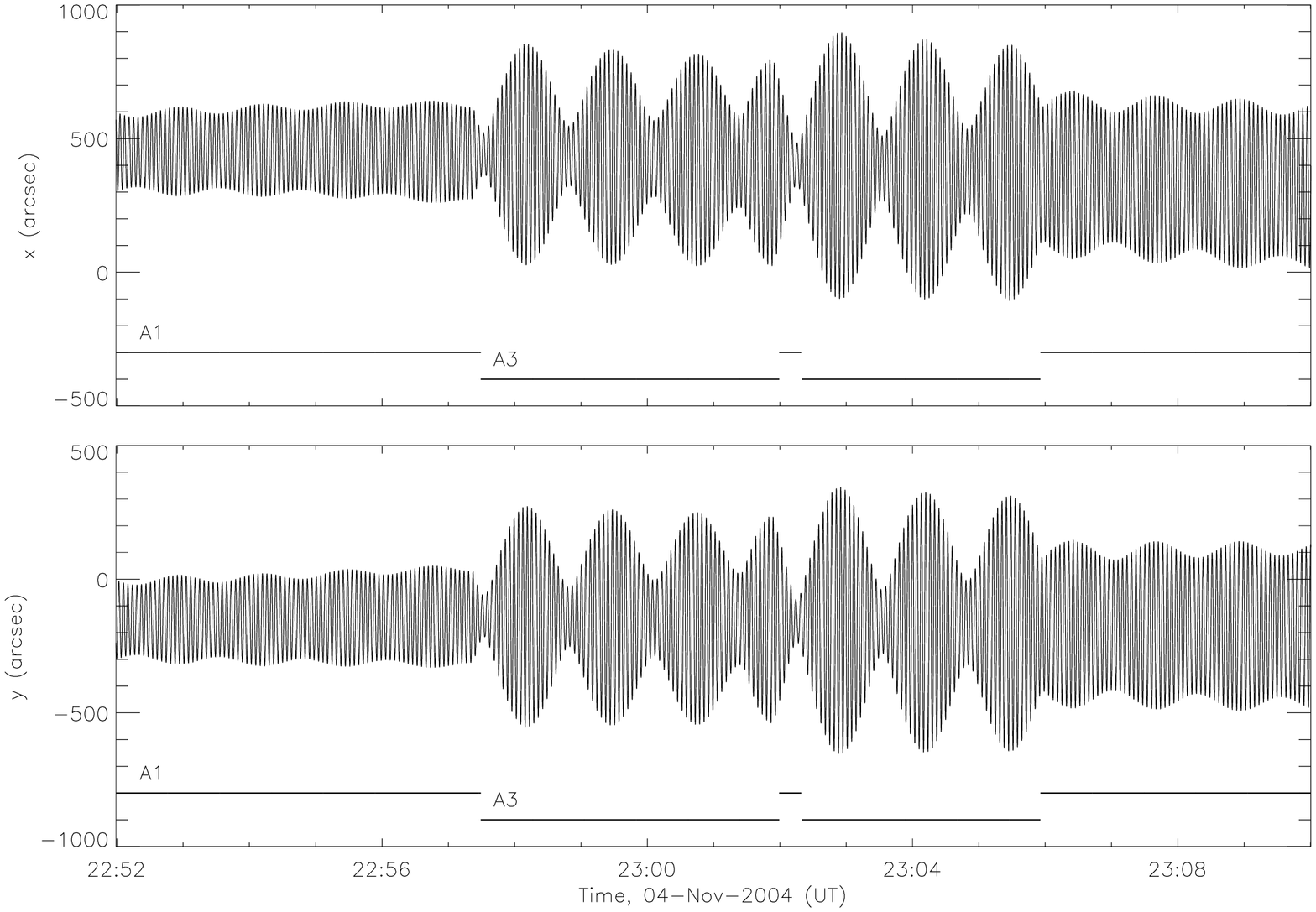}
   \includegraphics[width=6cm,angle=90]{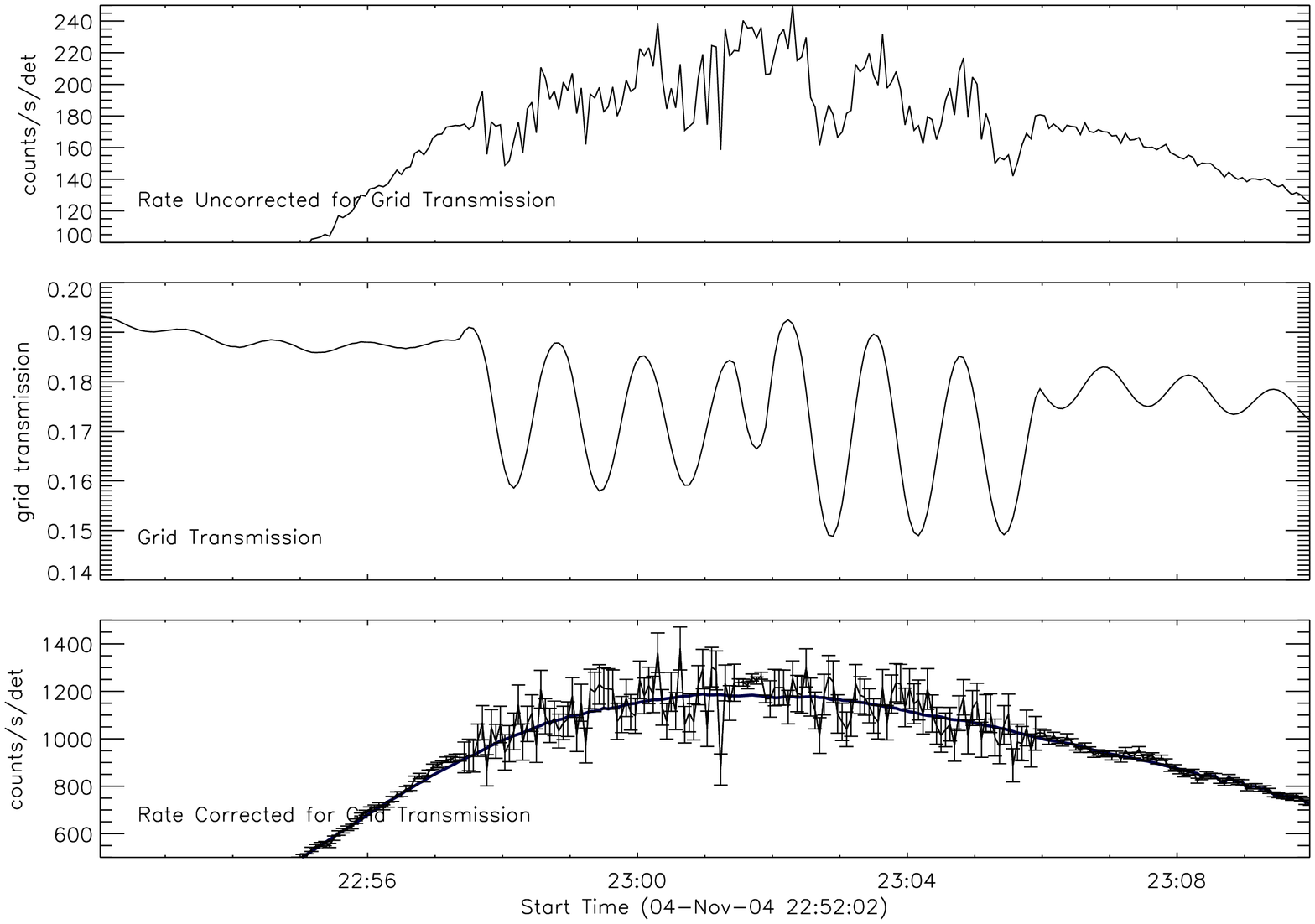}
\caption{Top: RHESSI imaging axis pointing variations in $x$ and $y$ during the impulsive phase of the 2004 November 4 flare. Changes in pointing modulation amplitude correspond to changes in attenuator state, from A1 to A3 at 22:57 UT, A3 to A1 and back to A3 at 23:02 UT, and finally A3 to A1 at 23:06 UT. Bottom: Uncorrected count rates detected in the 6-12 keV energy range by detector 5, grid transmission coefficient, and count rates corrected for grid transmission. The count rates in this figure have been scaled to account for attenuator state changes, hence the differing error bar sizes, which are based on Poisson statistics. The solid curve is the original data smoothed by a 50-point boxcar function.}
 \label{04_nov_04_impulsive_pointing}
\end{center}

\end{figure}

Figure \ref{04_nov_04_impulsive_pointing} shows that the A3 attenuator state is activated just after 22:57 UT. At this time the pointing variations of RHESSI increase markedly, until RHESSI reverts to the A1 state at 23:06 UT, where the pointing modulation returns to a lower level. As the bottom panel of Figure \ref{04_nov_04_impulsive_pointing} shows, 75 s oscillations can indeed be seen by detector 5 during this A3 interval, whereas they are not seen once RHESSI reverts to the A1 state.

This effect is key to understanding why RHESSI pointing varies between events, producing oscillations in some cases but not others. Since the attenuators are moving parts and RHESSI is spinning in free space, their activation affects the behaviour of RHESSI directly. The movement of the attenuators is enough to affect the pointing of RHESSI, leading to changes in nutation amplitude. The change in amplitude depends on the phase during rotation at which the attenuator motion occurs. This is the effect we observe in Figure \ref{04_nov_04_impulsive_pointing} and is a key contributor to the appearance of oscillations in RHESSI lightcurves. This effect also explains why the oscillations in the 2004 November 6 flare disappear at around 00:58 UT, when RHESSI returns to the A1 state.

\section{Concluding remarks}

From our analysis of the flares of 2004, November 4 and 2004, November 6, we can draw the following conclusions. Firstly, the observable oscillations in RHESSI X-ray count rates on 2004, November 6 are the direct result of large nutation motions of the satellite, which produce oscillatory motions of the imaging axis with respect to the spacecraft spin axis. The magnitude of the observed oscillations in the count rates varies considerably between detectors, as shown in Figure \ref{individual_detectors}. Detector 5 is an anomalous case and is disproportionately affected by imaging and spin axis motions. In the remaining detectors, the magnitude of the modulation is usually - though not always - small, and in some cases undetectable. This is caused by the misalignment of RHESSI's collimator axes with the spacecraft imaging axis. As measured prior to launch, each detector grid is subject to a slightly different offset from this axis, and in the case of detector 5 this offset is large - approximately 500 arcseconds. Hence this detector shows count rate modulation of greater amplitude.

Secondly, the extent of the modulation of the RHESSI imaging axis, and by extension the modulation in $\delta$, varies considerably between events. For example, during the 2004 November 6 flare the peak-to-valley change in $\delta$ was approximately 0.1 degrees, whereas during the 2004 November 4 flare the modulation was an order of magnitude smaller except when RHESSI entered the A3 state. Hence, in many RHESSI observations the effects of nutation are small or negligible. 

Hence, while RHESSI remains an effective tool for the study of QPP of X-ray emission in solar flares, as shown by the results of \citet{2005A&A...440L..59F, 2006ApJ...644L.149O, 2006A&A...460..865M, 2008ApJ...684.1433F, 2008SoPh..247...77L, 2009A&A...493..259I, 2009SoPh..258...69Z, 2010SoPh..263..163Z, 2010ApJ...708L..47N} for example, the fundamentals of its design mean that in some cases oscillations in count rates may arise as a result of its gyroscopic motions. We have shown this to be the case for the solar flare of 2004 November 6, where distinct oscillations of characteristic period 75 s were observed in the lightcurves during the decay phase of the flare and were initially erroneously interpreted as a magnetohydrodynamic process by \citet{ISI:000279654100009}. However, it is clear that this is an instrumental effect. Of particular importance is the offset between the RHESSI imaging axis and spin axis. This can vary significantly between events due to magnetic torquing, used by RHESSI to maintain its spin rate and to follow the Sun, and also due to motions associated with attenuator state changes. Large variations in pointing may, depending on the X-ray source position, result in significant changes in $\delta$ and hence lead to observable modulations in detector count rates. Detector 5 is particularly sensitive to this issue since its subcollimator axis has the greatest offset from RHESSI's imaging axis.

For the first time, we have implemented dynamical compensation for livetime and grid transmission in order to correct for nutation effects. This procedure should account for all the instrumental effects of the type discussed here, at least in the detectors with the finer grids, and should be used in all studies of this kind. Background subtraction and data gap correction for coarse grids must also be accounted for in future refinements. Also, the grid transmission correction incorporates pre-launch data, and thus it is possible that some residual effects may still persist and produce count rate modulations masquerading as QPP. Similarly, imaging and spectroscopy with RHESSI will be affected by the issues discussed in this paper. We intend that this solution will be refined and will become part of the RHESSI software.

%In conclusion, reliable detection of oscillations during solar flares remains challenging. RHESSI, with its high time cadence and wide coverage in the X-ray regime, remains one of %the most useful instruments in this regard. However, as we have shown here, it is important to be aware of the instrument's various motions and how they may sometimes produce %artificial oscillations. Hence, it is wise to remain skeptical during the study of such events, and verification by independent instruments remains extremely desirable.

\begin{acknowledgements}
We are grateful to Richard Schwartz for assistance with RHESSI data and software, and to Gordon Hurford for helpful comments. ARI was supported by an appointment to the NASA Postdoctoral Program at the Goddard Space Flight Center, administered by Oak Ridge Associated Universities through a contract with NASA. IVZ and ABS were partially supported by the Russian Foundation for Basic Research (grant No. 10-02-01285-a), by the grant NSch-3200.2010.2 and by the programm Minnauki Contract No. 14.740.11.0086. IVZ, VMN and ABS are also supported by the Royal Society British-Russian Research Collaboration grant.
\end{acknowledgements}

\bibliographystyle{aa}
\bibliography{qpprefs}

%\appendix
%\section{Dynamic correction for grid transmission and livetime}

%\begin{verbatim}

%obj = hsi_calib_eventlist()

%det_index_mask[4] = 1

%obj->set, det_index_mask = det_index_mask
%obj-> set, energy_band= [6.0000000D, 12.000000D]
%obj-> set, time_range=anytim('6-nov-2004 00:48:00')+[0,10*60.]
%obj-> set, time_bin_def= 128 + lonarr(9)
%obj-> set, time_bin_min= 1024L
%;Grid 5 data accumulated at freq of 1./64 HZ (2.^20/(16*1024.))
%obj-> set, use_auto_time_bin= 0L
%obj-> set, use_flux_var= 0L
%;Get the calibrated_eventlist structure as function of time for grid 5 (index 4)
%obj-> set, as_no_extrapol=0 ;trap bad aspect but don't crash out

%cbptr = obj->getdata()
%utp = obj->get(/ut_binned_eventlist) ;midpoints
%pvut=where(ptr_valid(utp))
%ut = get_edges(/mean, *utp[pvut[0]])
%nut = n_elements(ut)
%ngrid = total( ptr_valid(cbptr))
%cb0=reform(ptr_concat( cbptr,these_tag=['COUNT','LIVETIME','GRIDTRAN']),nut, ngrid)

%cgrate = f_div( cb0.count, cb0.livetime*cb0.gridtran)
%;set cb0.gridtran to 1 to show gridtran effects

%nsum = 32

%nutsm = nut/nsum * nsum
%livesm = total( reform( cb0.livetime, nsum, nut/nsum, ngrid), 1)
%uts = avg( reform( ut[0:nutsm-1], nsum, nut/nsum),0)
%cor_rate_grid =  f_div(  total( reform( cgrate * cb0.livetime,nsum, nut/nsum, ngrid), 1),livesm)
%cor_rate =  ngrid eq 1 ? cor_rate_grid : avg(cor_rate_grid,1)
%o=obj_new('utplot', uts-uts[0], cor_rate, utbase=anytim(/vms, uts[0]))
%o->plotman

%\end{verbatim}

\end{document}